\documentclass[12pt,a4paper]{article}
\usepackage{graphicx}
\date{}
\begin{document}

\date{}

\title{Gravitational waves signal analysis\\
{\normalsize The meaning of coordinates}}
\author{Ll. Bel\thanks{e-mail:  wtpbedil@lg.ehu.es}}

\maketitle

\begin{abstract}

I use a very simplified example to discuss the signature of a gravitational wave taking into account the relative state of motion of the detector with respect to the source that originated it. Something that to my knowledge has been ignored up to now and may ruin the best crafted template \cite{Templates}.

\end{abstract}

\section{Unperturbed asymptotic domain}

Let us start with Minkowski's space-time model:

\begin{equation}
\label{Minkowski}
dS^2=-dT^2+dX^2+dY^2+dZ^2
\end{equation}
so that the time-like congruence $S_0$ with parametric equations:

\begin{equation}
\label{S0}
X=X_0, \quad Y=Y_0, \quad Z=Z_0
\end{equation}
is a Galilean frame of reference.

The following coordinate transformation:
\begin{equation}
\label{CT}
Z=b^{-1}(e^{bz}\cosh(bt)-1), \ X=x, \ Y=y,\ T=b^{-1}e^{bz}\sinh(bt)
\end{equation}
leads to a second form of the line-element (\ref{Minkowski}):

\begin{equation}
\label{GF}
ds^2=-e^{2bz}dt^2+dx^2+dy^2+e^{2bz}dz^2
\end{equation}
and to a second frame of reference $S_1$ defined by the parametric equations:

\begin{equation}
\label{S1}
x=x_0, \quad y=y_0, \quad z=z_0
\end{equation}

The Jacobian of the coordinate transformation (\ref{CT}) is:
\begin{equation}
\label{Jacobian}
J=1, \ -\infty <\  z\ < \infty
\end{equation}
and therefore, although the two line-elements are covariant copies of one and another, the manifolds on which they are defined are not the same. It is $R^4$ in the first case and it has two borders in the second case.

The inverse coordinate transformation is:

\begin{equation}
\label{InvCT}
t=b^{-1}\tanh^{-1}\left(\frac{bT}{bZ+1}\right), \ x=X, \ y=Y,\ z=(2b)^{-1}\ln\left((1+bZ)^2-(bT)^2\right)
\end{equation}
Since the line-element (\ref{GF}) has been derived from (\ref{Minkowski}) by a coordinate transformation its Riemann tensor is zero.
\begin{equation}
\label{InvCT}
R^\alpha_{\ \beta\gamma\delta}=0
\end{equation}
However, if I consider the orthogonal decomposition:

\begin{equation}
\label{orthogonal}
ds^2=\eta_{\alpha\beta}\theta^\alpha_\mu\theta^\beta_\nu dx^\mu dx^\nu
\end{equation}
where:

\begin{equation}
\label{thetas}
\theta^0_\mu=-e^{bz}\delta^0_\mu, \ \theta^3_\mu=e^{bz}\delta^3_\mu, \ \theta^2_\mu=\delta^2_\mu, \ \theta^1_\mu=\delta^1_\mu,
\end{equation}
and its dual basis:

\begin{equation}
\label{invthetas}
e^\mu_0=-e^{-bz}\delta^\mu_0, \ e^\mu_1=e^{-bz}\delta^\mu_1, \ e^\mu_2=\delta^\mu_2, \ e^\mu_1=\delta^\mu_1,
\end{equation}
it follows that the Torsion of the corresponding Weitzenb\"{o}k connection:

\begin{equation}
\label{Torsion}
T^\alpha_{\beta\gamma}=-e^\alpha_\mu(\partial_\beta\theta^\mu_\gamma
-\partial_\gamma\theta^\mu_\beta)
\end{equation}
has one strict component that it is non zero. Namely:

\begin{equation}
\label{Strict}
T^0_{30}=b
\end{equation}

The space-time model (\ref{GF}) can be interpreted as describing an accelerated inertial frame of reference in the vertical $z$ direction whose geometrical origin is the covariant curvature of the time-like world-lines of the reference system:

\begin{equation}
\label{curvature}
\partial_zV=b, \ V=\ln(\sqrt{-g_{00}}), \ \ \ x^0=t
\end{equation}
or as a vertical gravitational field whose geometrical origin is Weitzenb\"{o}k torsion and not space-time Riemannian curvature \cite{Bel2}.

Other line-elements covariant equivalent to (\ref{GF}) with same time-like world-lines but different space-coordinates have been used. For example \cite{Moller}:

\begin{equation}
\label{Rindler}
ds^2=-(1+bz)dt^2+dx^2+dy^2+dz^2
\end{equation}
The line-element (\ref{GF}) I use is in agreement with my general theory of space-time models,\cite{Bel}, which requires to give a meaning to the coordinates of space by comparison with a flat 3-dimensional metric of reference. The model above complies with this theory in the sense that the coordinates $x,\ y\ z$ as $X,\ Y\ Z$ are harmonic coordinates:

\begin{equation}
\label{harmonic}
\triangle_4{x}=0,\ \cdots \ \triangle_4{X}=0,\ \cdots
\end{equation}
when the metric of reference is:

\begin{equation}
\label{harmonic}
d{\tilde s}^2=dx^2+dy^2+dz^2
\end{equation}

%%%%%%%%%%%%%%%%%%%%%%%%%%%%%%%%%%%%%%%%%%%%%%%%%%%%%%%%%%%%%%%%
\section{Frame dependence of retarded time}

Let us consider now a space-time model describing the asymptotic domain where a gravitational wave originated far away propagates. To our purpose it is sufficient to deal with the very simplified line-element:

\begin{equation}
\label{GW}
d{S^\prime}^2=-d{T}^2+(1+h(U))d{X}^2+(1-h(U))d{Y}^2+d{Z}^2,
\end{equation}
where the amplitude $h$ of the wave is a small function of the phase $U$:

\begin{equation}
\label{U}
 U=\frac{1-v}{\sqrt{1-v^2}}(T-Z)
\end{equation}
and $v$ takes takes into account a conditional relative velocity between the wave and a detector in the direction $z$ introducing a red-shift factor.

The space coordinates are again harmonic coordinates:

\begin{equation}
\label{Harmonic}
\triangle_4{X}=0+O(h^2),\ \cdots
\end{equation}
Since for $h=0$ the line element is the same as (\ref{Minkowski}), to take into account the acceleration of the detector, or equivalently the gravitational field in the area where it is located, I have to consider the transformation from one frame of reference to another described by the coordinates transformation (\ref{CT}) and therefore the gravitational wave line-element that has to be used is:

\begin{equation}
\label{GWprime}
d{s^\prime}^2=-e^{2bz}dt^2+(1+h(u))dx^2+(1-h(u))dy^2+e^{2bz}dz^2
\end{equation}
where now $u$ is:

\begin{equation}
\label{Signalu}
u=\frac{1-v}{\sqrt{1-v^2}}b^{-1}(e^{bz}(\sinh(bt)-\cosh(bt))+1))
\end{equation}
or, if $b$ is a small parameter:

\begin{equation}
\label{Signal0}
u=\frac{1-v}{\sqrt{1-v^2}}(t-z)\left(1+\frac12 b(t-z)\right)+O(b^2)
\end{equation}

\section*{Conclusion}

The simple example that I have worked out here shows the importance of taking into account i) the relative state of motion between the source of the gravitational radiation and its prospective detector, or ii) the local gravitational field in its environment. This requires to have a sound theory of reference that allows to use different coordinates having the same meaning in different frames of reference.

I propose that this work, conveniently developed, be integrated into the already complex craftsmanship of template banks that should help to detect gravitational waves.

%%%%%%%%%%%%%%%%%%%%%%%%%%%%%%%%%%%%%%%%%%%%%%%%%%%%%%%%%

\section*{Appendix}

Using the following notations:

\begin{equation}
\label{Notations}
b\cdot X=b_iX^i, \ b\cdot x=b_ix^i,\ |b|=\sqrt{b^ib_i}, \ b^i=\delta^{ij}b_j
\end{equation}
the generalization of the coordinate transformations (\ref{CT}) to refer Minkowski's space-time to a frame of reference whose world-lines have constant curvature $b_i$ is,

\begin{equation}
\label{CT3D}
x^i=(\delta^i_j-|b|^{-2}b^ib_j)X^j+(\sqrt{2}|b|)^{-2}b^i\ln((b\cdot X+1)^2-|b|^2T^2)
\end{equation}
It can be checked easily that:

\begin{equation}
\label{Dal}
\triangle_4 x^k\equiv-\frac{\partial^2x^k}{\partial T^2}+\delta^{ij}\frac{\partial^2x^k}{\partial X^i\partial X^j}=0
\end{equation}
meaning that the coordinates $x^i$ are harmonic.

The inverse transformations (\ref{InvCT}) become:

\begin{equation}
\label{InvCT3D}
X^i=(\delta^i_j-|b|^{-2}b^ib_j)x^j+|b|^{-1}b^i(e^{b\cdot x}\cosh(|b|t)-1), \ T=|b|^{-1}e^{b\cdot x}\sinh(|b|t)
\end{equation}
And  the transformed line-element is:

\begin{equation}
\label{potentials}
ds^2=-e^{2b\cdot x}dt^2+(\delta_{ij}+|b|^{-2}(e^{2b\cdot x}-1)b_ib_j)dx^idx^j
\end{equation}

\end{document}